# Emergence, causation and storytelling: condensed matter physics and the limitations of the human mind

Stephen J. Blundell

University of Oxford, Department of Physics, Parks Road, Oxford OX1 3PU, UK

*Abstract*

The physics of matter in the condensed state is concerned with problems in which the number of constituent particles is vastly greater than can be easily comprehended. The inherent physical limitations of the human mind are fundamental and restrict the way in which we can interact with and learn about the universe. This presents challenges for developing scientific explanations that are met by emergent narratives, concepts and arguments that have a non-trivial relationship to the underlying microphysics. By examining examples within condensed matter physics, and also from cellular automata, I show how such emergent narratives efficiently describe elements of reality.

## 1. Introduction

The subject of emergence has become of interest to philosophers (see O' Connor and Wong, 2015), and much has been written in aid of teasing out different types of emergent behaviour. A distinction is often made between `weak' and `strong' emergence, sometimes called `epistemological' or `ontological' emergence respectively, although different authors define these terms in slightly contrasting ways. Essentially, the distinction is between an emergence that is constructed in terms of the limits of human knowledge and one that is fundamentally irreducible, representing a new element of reality. For example, in the weakly emergent case, a macroscopic state can still be determined from the microscopic physics, but viably only through computer simulations that can crunch through repeated iteration of the low-level laws (Bedau, 1997). Thus it might be only difficult and cumbersome to go from the lower level explanation (the microscopic world) to the upper level (the macroscopic world), but not completely impossible. In the strongly emergent case on the other hand, the higher level is fundamentally irreducible to the lower level (Kim 1999), and new 'causal powers' are invoked which act 'downward'. Some philosophers seem to feel that the strong version is where the real philosophical meat is. Thus it is thought to be "the most interesting and important kind of emergence" (Silbertstein and McGeever 1999). Of course, such a `strong emergent' approach appears to best target their holy-grail problem, namely the determination of the nature of the conscious mind, which some wish to be wholly irreducible to physiological neural states (O'Connor and Wong 2005). Thus if I decide to act in the world, perhaps making up my mind to switch on an electric kettle, my strongly emergent consciousness (higher level) is imagined to downwardly cause a resultant state of (lower level) molecular motion in the heated water.

Strong emergence has however been described as "uncomfortably like magic" (Bedau, 1997) and most scientists have an instinctive aversion to it. Why shouldn't it be possible, in principle, even if not feasible in practice, to describe the entire process of me deciding to switch on a kettle and the resultant jiggling of $H_2O$ molecules all at the micro-level (neural and molecular processes) in a seamless whole? Scientists are perhaps expected to express such reductionist sentiments that would then predispose them against emergence, so the current popularity of the topic amongst physicists might be surprising. The word 'emergence' now frequently appears in the titles of research papers in condensed matter physics (in the last decade "emergence" or "emergent" has appeared in the title of well over a hundred papers in the journal *Physical Review Letters*). Some of the most vocal advocates of emergence have been condensed matter physicists (Anderson 1972, Laughlin and Pines 2000, Laughlin 2005). The Oxford English Dictionary defines emergence as the "process of coming forth, issuing from concealment", and this appearance or manifestation of something that was previously buried or hidden from view captures the sense of the word as used by physicists. Emergent properties are somehow inherent in the underlying microscopics, but not in any obvious or easily extractable manner, and their appearance is wonderful, surprising and pointing to higher-level organizing principles that operate at a new level. But are these higher-level organizing principles simply weak emergence?

Some string theorists in fact reject the idea that emergent principles represent new physics at all, even though they might be important for practical purposes. Brian Greene states that although "it would be hard to explain the properties of a tornado in terms of the physics of electrons and quarks, I see this as a matter of calculational impasse, not an indicator of the need for new physical laws. But again, there are some who disagree with this view." (Greene 2000) So are emergent laws new or do they simply represent what Greene calls a "calculational impasse"?

The idea that I will develop in this paper is that emergent laws and properties are independent, novel structures that function effectively because they are well adapted for human thought processes. I discern that there is a powerful analogy between doing science and storytelling. To work, stories have to be succinct, told well, have a point and express some truth. This is simply because they are transmitted and received by human minds which have certain physical limitations. These limitations become crucial when faced with any physical problem involving complexity, be that a story of human interactions (such as *Middlemarch*) or a story of electronic interactions (such as the Mott insulator). Emergent properties nevertheless have both an ontological and an epistemological character. To develop this thesis I will begin in Section 2 by considering some lessons that can be extracted from Conway's game of *Life* and the insights that it gives on the nature of causation and the nature of what I call emergent narratives. In Section 3, I will develop the idea that emergent narratives are effective because of the physical limitations that apply to human minds and will describe these limitations by analogy with Landauer's notion of the physicality of information. In Section 4, I will give examples of how these emergent narratives can be successful in scientific descriptions of systems in condensed matter physics.

**2. Life Lessons**

John Horton Conway's game of *Life* is a favourite example of a simple mathematical system (a cellular automaton) that illustrates surprising and unexpected complex behaviour and was introduced as a toy model for understanding the appearance of biological organisms. It is still a useful starting point for describing emergent properties. The principles of the game of *Life* (Gardner 1970, Poundstone 1985, Adamatsky 2010) are simply stated: A square grid of cells (like a checkerboard) is defined and each of its cells can be either alive or dead (and usually coloured black or white respectively). Time is discretized and a new configuration is obtained at each time-step that is determined only by the configuration during the previous time step, and worked out according to the following four rules.
1. Any live cell with fewer than two live neighbours dies, as if caused by under-population.
2. Any live cell with two or three live neighbours lives on to the next generation.
3. Any live cell with more than three live neighbours dies, as if by overcrowding.
4. Any dead cell with exactly three live neighbours becomes a live cell, as if by reproduction.

These simple rules are all there is, and so we have stated the `theory of everything' for the *Life* universe (just in fact as we do in condensed matter physics, where the theory of everything is the many-particle Schrödinger equation, see Laughlin and Pines (2000)). Thus *Life* is ideal for discussing emergence (Bedau, 1997). Of course, the game of *Life* has clear limitations for describing `our' world: it's only two-dimensional, the state space is 1 or 0, space is quantized, evolution is irreversible (the arrow of time is hard-wired in) and it's completely deterministic. Nevertheless, as we shall see it is highly illustrative of many features of the real world.

The game of *Life* provides a privileged standpoint, just as imagined for an all-seeing Laplacian demon (of which more later). One is able to observe the entire *Life* universe, staked out on its grid, and follow its evolution in minute detail as time iterates forward. A physicist would naturally look for stable structures in the game of *Life*, and in fact one quickly finds that there are several (these are known as `still lifes'). But then you encounter `oscillators', strange forms that loop periodically through a sequence of configurations, some simple (the `blinker' is a period 2 oscillator consisting of three squares in a line), some highly complex (the `queen bee shuttle' is a period 30 oscillator and some are known with periods of several hundred). These `life forms' are all rooted to the spot, but there are also `spaceships', configurations that propagate across the grid. *Life* has a natural speed limit since the rules dictate that each cell can only influence its direct nearest neighbours, so that the effective speed of light, $c$, (the speed at which some effect can propagate) is one cell per generation (though because of *Life*'s nearest-neighbour rule, this can be either along the horizontal/vertical axes or diagonally between them and so is rather anisotropic). A spaceship must retain its shape while propagating, and Conway showed that this means that spaceships can travel horizontally or vertically no faster than $c/2$ (spaceships are known that travel at $c/2$, $c/3$, $c/4$, $17c/45$, $31c/240$ and many other values, but $c/2$ is the upper limit). Diagonal transport is also possible; an example is shown in Figure 1(a) and looks more like a flying bird than a spaceship and accordingly is known as a `Canada goose'. It flies diagonally (Figure 1(b)), at a speed of $c/4$ (after four generations it will have advanced one cell horizontally and one cell vertically) and if you watch an animation of its movement it rather resembles a bird in flight, gently flapping its wings. Its detailed structure is quite critical to its operation. Removing a single pixel from the initial configuration (Figure 1(c)) results in the bird exploding in what looks like a fireball only shortly after takeoff (Figure 1(d)). As with some mutations of DNA, or tiny chemical changes on a small molecule, the effects of a minor alteration can have dramatic consequences. (In the *Life* world the forms exhibit much greater fragility than those in the physical world. In our world, the higher spatial dimension, the greater complexity of the underlying laws and the rigidity arising from broken symmetry states all contribute to a more robust stability against perturbations.)

Although the simplest structures in *Life* can be derived using pen and paper (as Conway himself did), more complex structures require a computer to mindlessly, but accurately, iterate the rules. In one sense, *Life* is simply a grid of ones and zeros twinkling in and out of existence in the unthinking service of an unrelenting algorithm. So is the Canada goose real? Or is this just a pattern I see, telling you more about my brain than about the nature of `reality'? Since the Canada goose is a well-defined structure in *Life*, with measurable properties such as shape, speed and direction of travel, it is as real as any of the pixels out of which it is composed (see Dennett 1991). In the physical world, we describe particles as excitations in a quantum field. That field pervades all space, but a particle `exists' when that field is promoted out of the vacuum state at some position in space (see e.g. Lancaster and Blundell 2014). The particle has well defined properties such as mass and charge (and, for the photon, a fixed speed and direction of travel, just like the Canada goose). So these structures are real, but have to be perceived at a higher level of description from that of the individual particles, and this is what emergence is all about.

More complex structures are still being discovered in *Life* (see www.conwaylife.com/wiki for more details) and are found using the intelligence of real people, creatively using meta-rules about important *Life* processes that go beyond the basic rules. To give an example, in language that I have designed to look like physics, consider the collision of two horizontally travelling spaceships (Figure 2); as is common in physics, one tries to understand systems by smashing them into each other. The results reveal a plethora of interesting phenomena, but I have chosen just three examples. By varying the initial positions (and in the final case using a slightly longer spaceship) the results of the collision are seen to be quite different, leading to the creation of two `gliders' travelling diagonally (Figure 2(a,b)), total annihilation (Figure 2(d,e)) or creation of a `pulsar', a visually attractive period 3 oscillator (Figure 2(g,h)). It is easy to describe the `before' and `after' of each collision, but the process of the collision itself is quite complicated, so I have written down a kind of Feynman-like diagram to conceptualize the interaction (Figure 2(c,f,i) show diagrams for these three processes). These diagrams are clearly much easier to comprehend, and it is this kind of modular insight (which one could call an 'interacting field theory of *Life* structures'), focussing on the function of larger structures, that has led to the construction of logic gates, information processors and Turing machines within the game of *Life* (Rendell 2002, Rennard 2002).

Note that the causal flow in both levels, lower (individual pixels) and upper (Canada geese), are independent and you can use either (see Figure 3). It might be convenient to stick to the upper level when things are simple and regular, such as spaceships flying through the air in straight lines, but then dive down to the lower level to compute the collision process, and then rise back up to the upper level afterwards. This probably provides the cleanest explanatory account, but note that the swapping between levels is simply *your* choice and so the apparent top-down or bottom-up causation in those vertical jumps in Figure 3 merely trace out the causal path *you* have selected. It is entirely legitimate to fix your whole attention either on the lower level (as the computer does) or on the upper level (invoking a Feynman-like procedure to handle the collision process). The switching between levels depends on how you want to think about the physical processes and is therefore purely epistemological. (In this, I share Butterfield's unease (Butterfield, 2011) unease about many discussions of top-down causation that reify one particular direction of causal flow.) Nevertheless, there is an ontological dimension to the levels themselves. The 'real patterns' (Dennett 1991) at the higher level deserve ontic status every bit as much as do the flickering pixels at the lower level. Both levels are valid descriptors of the `reality' of properties within the *Life* universe, but just as we find in the physical universe, certain levels are better suited to brute computation, others to the construction of narratives comprehensible to the human mind.

Even though the `theory of everything' is known, we have found that new structures and new `laws' emerge non-trivially and unexpectedly. It is often stated that the game of *Life* is a good example of `weak emergence', weak because unlike the case of `strong emergence' you can (it seems) always compute everything at the lower level. Thus reductionism works, and you don't need to work at the higher, emergent level if you don't want to. Of course, *Life* is an example where you can sit back and let the computer take the strain and work everything out. But the computer doesn't pull the patterns out for you as it computes at the level of the flickering pixels; it needs you, the observer, to *see* the Canada geese. And the computer only simulates the microscopic world because the grid we choose is usually very small; for the game of *Life* implemented on a grid of size $10^{12}$ pixels by $10^{12}$ pixels then one single configuration would exceed the total storage capacity of all computers currently on Earth (and if this paper is being read in the far future, increase the dimension of the grid by a few powers of ten until my argument holds). Thus we can't simulate on these scales. Nevertheless, with only a modicum of thought, you can work out how many time steps it would take for a Canada goose to fly from one corner to the other using an emergent meta–law (Answer: $4 \times 10^{12}$).

The game of *Life* is often dismissed as just weak emergence since, without anything complicated happening, the simple rules allow the automata to iterate along in a manner that is calculable (but only, I stress, for relatively small systems). But the `biologists' of *Life* have spent decades studying the taxonomy of these 'Life-forms' and the major discoveries that have been made required radically emergent thinking and a deep and profound "knowing" of the problem that goes way beyond mere simulation. Thus the power of these emergent laws should not be underestimated (it is not simply, in Greene's phrase, a matter of "computational impasse"). With a Turing machine constructed within the game of *Life*, it is possible to construct an initial condition for which the final result is genuinely undecidable (Moore 1990, Bennett 1990, Wolfram 1985) so that even this simple `game' belies the presence of extraordinarily subtle behaviour.

A more complex example is that of number theory. The 'universe' of number theory is the set of integers, equipped with the basic 'low-level' rules of addition and subtraction. Yet the richness of the structure, patterns and forms latent in this apparently bland arithmetic structure have for centuries dazzled and baffled some of the finest mathematical minds. The emergent laws are, as in the game of *Life*, reducible to the basic rules of the system, but once again the way they emerge is highly non-trivial and requires the development of new emergent concepts. There is a fundamental difference between being able to calculate within a system and knowing it at a deeper (higher) level. As Wigner has put it, mathematics "would soon run out of interesting theorems if these had to be formulated in terms of the concepts which already appear in the axioms" (Wigner 1960).

*Life* and number theory are simply defined arenas in which rich emergent behaviour nevertheless unfolds. The physical world is equipped with an even more complex mixture of competing interactions and varieties of particles even at the lower microscopic level, so that one can expect the emergent higher level properties to be even more startling and abundant. I note that Silberstein and McGeever in their characterization of different types of emergence focus on entanglement of identical quantum particles as a good example of ontological emergence. Entanglement demonstrates a failure of whole-part reductionism, so that an entangled pair "gives us good reason to doubt the atomistic vision of the world" in which "fundamental particles carry for ever fixed properties independently of their contextual features" (Silberstein and McGeever, 1999). Of course, the quantum mechanics of an entirely empty Universe containing a single particle is a barren, sparse, scrawny theory. Quantum mechanics demands a more abundant and lush landscape to display its richness, and an entirely empty Universe containing two particles is the absolute minimum requirement! But their emphasis on contextual features correctly highlights the potential for the relationship between entities to generate new features of reality. These new emergent features, such as quantum entanglement, fully deserve their ontic status, but I will argue that this is in common with most, if not all, such emergent features.

## 3. The limits to knowing

What does it mean for us to get our head around a physical system? In a two-body problem, it is possible for our minds to keep track of the positions and momenta of the two bodies, though we frequently work in a reference frame where we keep one of the two fixed (we speak of the Earth going round the Sun more often than the two orbiting their centre of mass). By keeping track of these variables, I can make a one-to-one correspondence between the value taken by a physical variable at a particular time and a number which is stored in my brain, or written on a piece of paper, or stored in a computer (if I wish to let a machine take the strain). With the values taken by those variables I can predict future behaviour or retrodict past behaviour using relatively simple analytical formulae. The three-body problem is vastly more complicated and resists simple analytic description in most cases, but the physics of matter in the condensed state frequently involves a $10^{23}$-body problem. In this case, qualitatively new behaviour emerges; the existence of more bodies is not just a simple change of scale but in Anderson's memorable phrase ``more is different'' (Anderson, 1972). $10^{23}$ vastly exceeds the number of things a human can conveniently think about (we each have fewer than $10^{12}$ neurons and $10^{15}$ synapses, but most of us can only focus on a half-dozen objects at one time). Moreover in these problems we frequently need to think about combinatorial numbers such as $10^{23}!$, a number that is larger than ten to the power of $10^{24}$. Such a number vastly exceeds the number of particles in the observable universe, and thus there is insufficient physical computing resource to calculate in a one-to-one sense.

The revolution in our understanding of information science is neatly encapsulated in Rolf Landauer's aphorism: "Information is physical" (Landauer 1961, Bennett 2003). Computer science had been thought to operate in an entirely separate domain from the physical world, an abstract space of ones and zeros interacting via chains of logic gates and churning through algorithms, but entirely divorced from the physical world. Landauer's insight was to see that any string of information has to have a physical embodiment, whether written down on a piece of paper, stored as charges on the gates of transistors in a chip, or held within a human mind. Thus even information is subject to the laws of physics (Parondo et al. 2015). This idea led to the resolution of the paradox of Maxwell's demon (reviewed in Leff and Rex 2003), the imaginary intelligent agent that, by opening and closing a small shutter connecting two volumes, could sort out dissimilar molecules and apparently circumvent the second law of thermodynamics, effortlessly bringing order and decreasing the entropy of the Universe. But the demon performs an elementary computation to sort each molecule, and must use at least one bit of storage in the process. Though the computation can be carried out reversibly, the process of memory erasure is irreversible. Thus, the demon either accumulates a larger and larger record of its past computations (quickly exceeding its physical memory allocation, since it would need an Avogadro number of bits for each mole of gas sorted) or it resets its memory, erasing bits and causing heat dissipation whose net result precisely cancels any entropy reduced.

In my view, such an approach also has consequences for the Laplacian demon, the imaginary agent that can supposedly view a physical system and know it in its entirety, without the tiresomely limited view afforded to an experimental physicist. If such a demon were (even hypothetically) to be constructed in our physical world, it would be subject to physical constraints which would include a limit on the number of atoms it could contain, bounded from above by the number of particles in the observable Universe (see also Lloyd 2002). Hence there is insufficient physical resource in the entire Universe to allow for the operation of a Laplacian demon able to analyse even a relatively limited macroscopic physical system. I suggest that these physical limits of knowability affect not only physicists, but also philosophers. These physical limits also provide constraints on what one can really say meaningfully about ontology, the nature of reality. The "view from nowhere" (Nagel 1986) that appears to be implicitly (and sometimes explicitly) championed by some philosophers is inconsistent and entirely untenable for observers, including philosophers, who are physically embodied in the Universe.

All of this is not to say that reality is unimportant, but physicists and philosophers alike perceive it and make statements about it on the basis of limited knowledge and partial perception, so ontological statements always have an epistemological dimension. This is where emergent explanations or narratives are ideal because they make a snug fit with the manner in which our minds are constituted. It is the nature of these emergent explanations that I will now consider.

**4. Emergent narratives**

A narrative in a history or a novel charts a comprehensible path through a morass of human complexity and interactions, helping the reader to focus attention on key events and the links between them. Our minds cannot cope with the Laplacian-demon viewpoint of a human drama, knowing every single event and character in excruciating detail. The author makes choices, and their skill often lies in what to leave out rather than what is included. The resulting narrative provides enough structure to capture the essence of reality without bogging down a finite mind with unnecessary and inconsequential detail, and an economy of style and expression in a narrative is frequently praised.

It seems to me that emergent theories and explanations function in similarly in the scientific domain, with emergent narratives capturing the essence of reality in a way that is far better fitted to the constraints and preferences of the human mind that a brute description of all the details at the lowest level. My intention is not to evaluate to what extent particular examples of scientific literature display characteristic features of narrative construction (as has been done elsewhere, see e.g. Norris et al. 2005). Neither am I concerned with the notion of "models as fiction" which alleges that by employing idealizations and abstractions scientists engage in a type of "make-believe when they use nonrealistic descriptions to model phenomena" (this view is discussed in Morrison 2015, from which this quote is taken). My purpose is different and rests on the assertion that any scientific discourse that aims to promote understanding has a narrative essence because it has to tell a story of complexity to a finite mind.

Condensed matter physics provides some good examples of this. For example, one strategy for coping with the challenge of 'knowing' something about macroscopic systems is to identify the right quasiparticle. As a simple example, consider a semiconductor, a material such as silicon, in which an energy gap (known as the band gap) separates the valence band and the conduction band. At absolute zero, the valence band is completely filled with electrons and the conduction band is completely empty. As the temperature increases, it becomes possible to promote a few electrons from the valence band to the conduction band, and these electrons become mobile in the conduction band and hence conduct. However, in the valence band there are now a few empty states, known as holes, and these too can become mobile. What does that mean? When an empty state (a hole) moves one jump to the right, it is *really* an electron that moves one jump to the left. But the concept of a hole is useful because we focus on a few holes rather than the huge number of electrons. The hole has some strange properties (such as having a negative mass), but the price paid for this modest imaginative investment is outweighed by the usefulness of the 'hole' concept. (In much the same way, we may worry about a bubble rising in a glass of beer since gravity should pull it downwards, but of course the heavier liquid flows around it – but we focus naturally on the rising bubble and not the falling liquid.)

Physics is replete with many other examples of these emergent phenomena. For example, understanding the thermal properties of solids is accomplished using 'fictitious' quasiparticles called phonons, which are the vibrations of the crystalline lattice whose energy is available in quantized lumps. These are collective modes of the atoms in a crystal, but they behave like particles; you can bounce neutrons off them and measure their energy-momentum relationship (their dispersion relation), just like any other particle. They behave just like real particles, and are excitations in the phonon field just as electrons are excitations in the electron field. The same goes for magnons (quantized spin waves), plasmons (quantized plasma waves) and a host of other examples (Anderson 1984), all of which qualify as emergent particles equipped with ontic status, each being described by the same type of field theory as is used for 'fundamental' particles such as electrons and photons.

In each of the examples discussed above, a classical harmonic mode succumbs to 'second quantization' (the quantum mechanical appearance of discrete particle-like structure out of a classical wavelike model) giving rise to emergent particles, but I now present a recent example where the emergent particles have a quite different origin. A particular magnetic crystal, $Dy_2Ti_2O_7$, has a crystal structure in which the dysprosium (Dy) ions are arranged in a network of corner-sharing tetrahedra (for our present discussion, we can forget about the other ions, Ti and O). Each dysprosium ion sits at the corner joining two adjacent tetrahedral. The dysprosium ions are magnetic and the crystal field (the electrostatic effects on neighbouring ions acting on the magnetic energy levels) constrains the magnetic moment (known as a *spin* for short) of each dysprosium ion to point along the axis joining the centre of the two adjacent tetrahedra, out of one and into the other (this results in classical Ising-like behaviour). When you include the magnetic interactions between the dysprosium ions on the network of tetrahedra, you find that the rule of the game is now that two of the spins can point in and two of them can point out. It doesn't matter which two are in, and which two are out, but the rule: '2-in, 2-out' has to be followed. When you extend this throughout the whole crystal, the freedom to choose which spins are pointing in and which are pointing out gives an additional entropy to the system – a residual disorder which persists to low temperature – and this can be measured in experiments. It turns out that the statistical mechanics describing this situation are entirely analogous to that of proton disorder in (water) ice, and so this compound is known as *spin ice* (Harris et al. 1997). An example of a spin ice configuration is shown in Figure 4(a) for the simpler case of a two-dimensional lattice of corner-sharing squares. Here each spin belongs to two squares and each square satisfies the '2-in 2-out' rule.

The underlying '2-in 2-out' dictat (and which is known in the trade as the 'ice rule') results in a divergence-free magnetization. It has now been appreciated that this leads to the appearance of an emergent form of electromagnetism, namely a description of the system in terms of an emergent gauge field that reproduces aspects of conventional electromagnetism that supports topological excitations (Castelnovo et al. 2008). To explain what this means, let us ask what happens if we put a mistake into the structure? What if we reverse a single spin? In this case, one of the tetrahedra will have '3-in 1-out' (let's call this configuration +) and, because the tetrahedra are corner sharing, a neighbouring tetrahedron will have '1-in 3-out' (let's call this –). This situation is illustrated for the two-dimensional lattice in Figure 4(b).

The key insight is to appreciate that this second tetrahedron can be restored to its ideal '2-in 2-out' state by flipping a magnetic moment on its other side. What this does is to shift the '1-in 3-out' configuration along. We can then repeat the trick and shift the '1-in 3-out' configuration further away from the '3-in 1-out', so that these two rule-breaking configurations can each move independently through the spin ice (see Figure 4(c)). Essentially we have 'fractionalized' the reverse spin, breaking it into two (the + and -) and allowing them to separate and go their own way. In fact, it turns out that the separated `halves' of the magnetic moment behave like individual *magnetic monopoles* (Castelnovo et al., 2008). (Note that Maxwell's equation, div ***B***=0, is not violated, as these monopoles represent particles for which div ***H***≠0.)

Now the magnetic monopoles in spin ice are, at root, composed of 'nothing but' atomic magnetic moments, obeying Maxwell's equations. However, the most efficient description of the phenomenon is obtained by describing the system in terms of quasiparticles, which in this case are magnetic monopoles. Hence we can subtract the background 'vacuum' state of spins away from the problem and focus only on the monopoles. This is a radical strategy because the vacuum here is a very rich structure of spins populating the lattice in a divergence-free (spin ice) configuration. But subtracting it away gives rise to a dramatic simplification that gets to the heart of the key physics. Thus considering the physical situation with only the monopoles (Figure 4(d)) is far simpler than if our attention is purely on the spins (Figure 4(c), and imagine that diagram without the monopoles and their path through the lattice so clearly indicated).

But isn't this just weak emergence? Are not scientists simply struggling with their imperfect models and wrestling with questions of epistemology, rather than addressing reality head on? I reject such a clear-cut distinction. Emergent properties are members of the set of elements of reality, and as such merit ontic status. Moreover, human minds have fundamental limits imposed on them by the physical nature of the universe, and though we (physicists and philosophers) aim at making firm statements about reality in order to construct a coherent ontology, our viewpoint is from within that universe, not from outside it, and is consequently constrained. Even in thermodynamics (how more real can you get?) our fundamental notion of entropy is one that depends on the information accessible to the experimenter and its limits (Jaynes 1957). That limit of our knowledge mandates that ontology can never be performed 'in a vacuum', viewed 'from nowhere' without some measure of epistemology that takes into account our own participation in the Universe. Moreover, not every emergent narrative will correctly pick out a set of elements of reality, and those narratives that are totally misguided or even very slightly flawed have to be weeded out or adapted, however imperfectly, in a process driven by new experimental and theoretical developments.

My approach is contrary to the claim that "epistemological emergence does not have any obvious ontological implications; but ontological emergence does" (Silberstein and McGeever 1999). The emergent structures I have been discussing are derivable from a lower level, but all have ontological implications. They also have, as discussed above, an unavoidable epistemological component. However, there are indeed distinctions that can be made between emergent structures according to how easily derivable they are from lower level descriptions (and could provide a continuum that one could label as ranging from 'weak' to 'strong'). At one end of this spectrum there is the concept of angular momentum: you might only need Newton's laws when computing a simulation of galaxy formation from dust, as your supercomputer computes the forces and crunches the dynamical laws for a large number of gravitationally attracting particles, but the emergence of angular momentum and its conservation (inherent but not explicit in Newton's laws) greatly simplify the story you tell of why the galaxy in the simulation comes out to be a spiral shape. Angular momentum as an emergent property (not put in `by hand' at the start or immediately obvious from staring at the force laws or equations of motion) is of course reasonably easy to derive, but there are plenty in many-body quantum mechanics that are not, and so may be located further out on the spectrum. But all these emergent structures have an ontic status and to dismiss them as merely epistemic devices seems to miss the point.

## 5. Conclusion

The restrictions of the human mind force the following behaviour for any non-trivial phenomenon: scientists select a bit of the Universe for study, and decide to focus on what they discern to be the key aspects, naturally locating the key emergent properties. Following any understanding that they glean, a story (explanation) is written. The best storytellers will find the right language for the story – perhaps a mixture of words, mathematics and pictures – and the search for the best story is a highly non-trivial process. Stories can be pictorial, as in Feynman diagrams (used above for collisions in the *Life* universe and in real life for quantum electrodynamics) which function as a kind of comic strip narrative. This process helps those of us `hearing' the story to `see' the point. Because of human limitations, both in the tellers and hearers, unnecessary details are left out (and that involves a degree of choice which may turn out to be judicious or foolhardy – in some situations there are principled reasons for ignoring the details, see Berry 1994, Batterman 2001). Moreover, an emergent explanation is not a simulation so will point to reality, even if it is not in one-to-one correspondence with it. Crucially, though finite an emergent story can speak of the infinite. Thus these emergent narratives are how in reality we navigate a complex world.

I would like to acknowledge useful conversations with Bob Batterman, Katherine Blundell, Harvey Brown, Alex Carruth, Lorenzo Greco, Tom Lancaster, Tim O'Connor, Mark Pexton, Wilson Poon, Simon Saunders, Christopher Timpson and David Wallace and also to thank the Durham Emergence project for their hospitality at their summer conferences, during which some of these ideas were developed.

Nagel, T. (1986) *The View from Nowhere*, Oxford University Press, Oxford.

Norris, S. P., Guilbert, S. M., Smith, M. L., Hakimelahi, S., and Phillips, L. M. (2005) "A theoretical framework for narrative explanation in science" *Science Education* 89, pp. 535-563

O'Connor, T., and Wong, Hong Yu (2005). 'The Metaphysics of Emergence', *Noûs*, 39, pp. 658-678.

O'Connor, Timothy and Wong, Hong Yu, "Emergent Properties", *The Stanford Encyclopedia of Philosophy* (Summer 2015 Edition), Edward N. Zalta (ed.).

Parondo, J. M. R., Horowitz, J. M. and Sagawa, T. (2015), "Thermodynamics of information", *Nature Physics* 11, pp. 131-139.

Poundstone, W. (1985) *The Recursive Universe*, Dover, Mineola, New York.

Rendell, P. (2002) ``Turing Universality of the Game of Life'', in A. Adamatsky (Ed.), *Collision-Based Computing*, (pp. 513-539) Springer, London.

Rennard, J.-P. (2002) `` Implementation of Logical Functions in the Game of Life'' in A. Adamatsky (Ed.), *Collision-Based Computing*, (pp. 491-512) Springer, London.

Silberstein, M., McGeever, J (1999) ``The search for ontological emergence'', *Philos. Q.* 49, pp. 182–200.

Wigner, E. (1960) "The unreasonable effectiveness of mathematics in the natural sciences", *Communications of Pure and Applied Mathematics* 13, pp. 1-14.

Wolfram, S. (1985) "Undecidability and intractability in theoretical physics", *Phys. Rev. Lett.* 54, pp. 735-738.

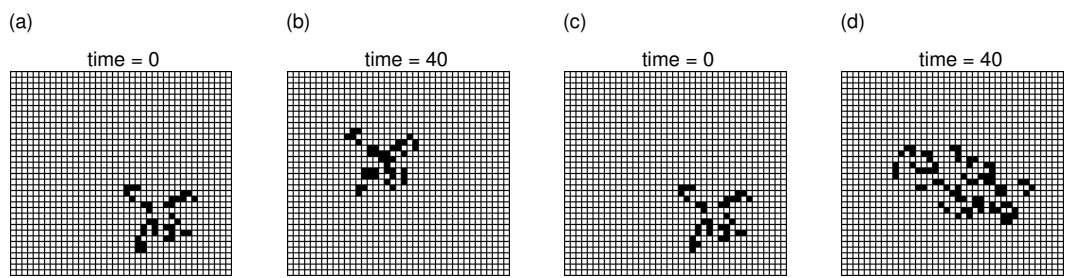

Figure 1: The Canada goose structure in the game of *Life* flies diagonally [shown in (a) and (b) forty time-steps apart]. (c) Changing one pixel in this structure results in the canada goose disintegrating after several time-steps [shown in (d)].

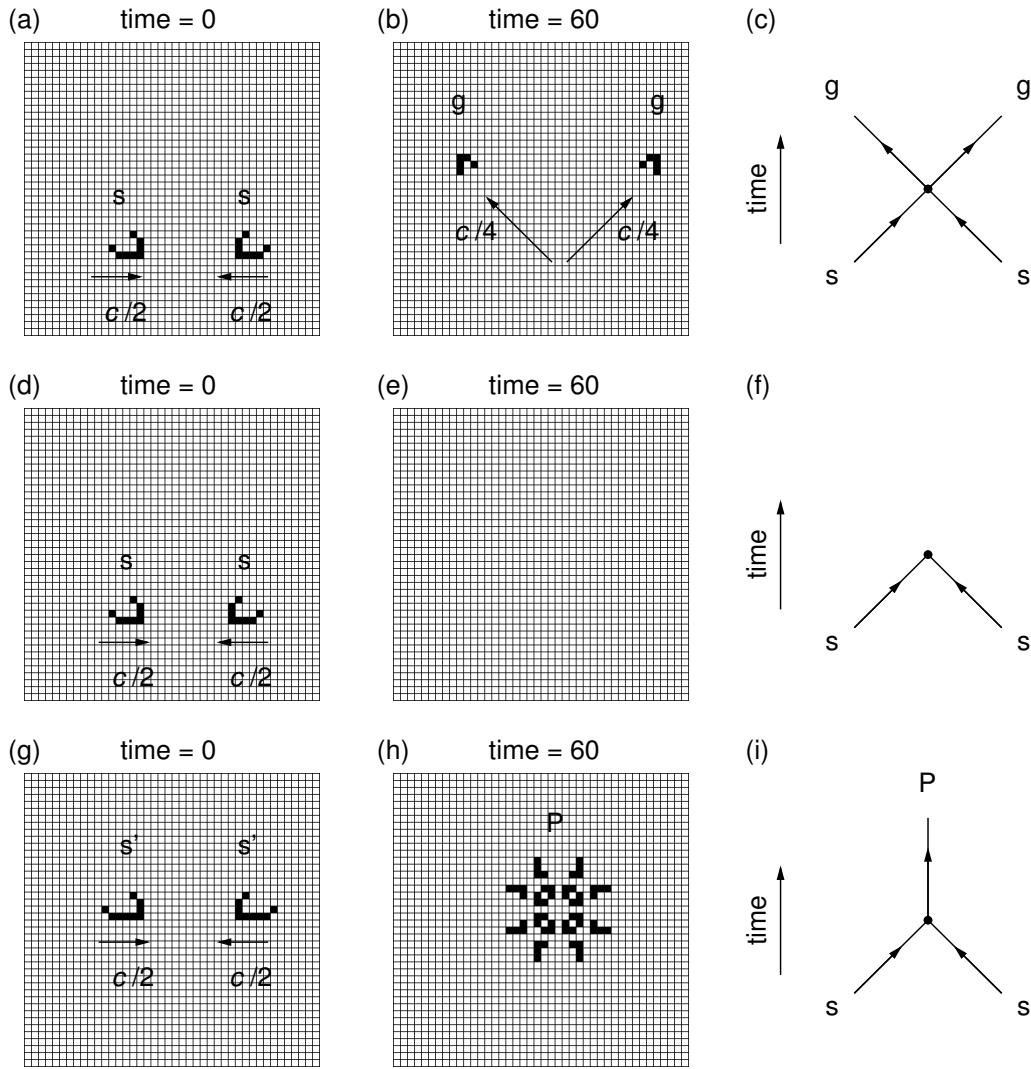

Figure 2: The collision of two horizontally travelling spaceships (with speed c/2). Depending on the initial conditions, this can lead to the creation of two 'gliders travelling diagonally with speed c/4 (a,b), total annihilation (d,e) or creation of a structure known as a pulsar, a visually attractive period 3 oscillator (g,h). These processes can be described succinctly using a Feynman-like diagram to conceptualize the interaction (c,f,i). (Here 's' represents a spaceship, 'g' a glider and 'P' a pulsar.)

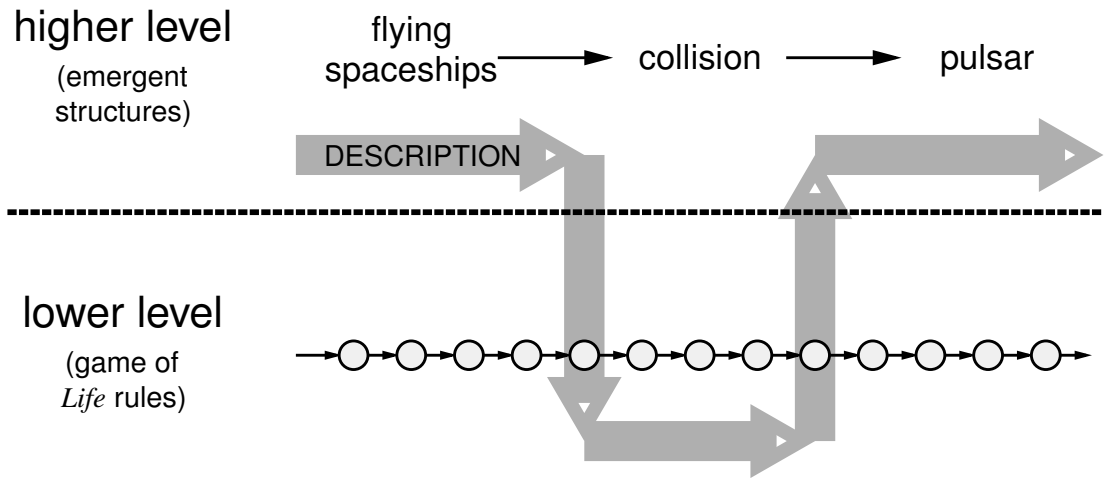

Figure 3: Processes in the game of *Life* can be viewed at a lower level, iterating the rules, or at a higher level, focussing on the emergent structures. A best description of the process in Figure 2(g,h,i) could stay at the higher level until something complicated occurs (a collision), when attention dips down into the lower level, before rising back to the higher level when simple behaviour reappears.

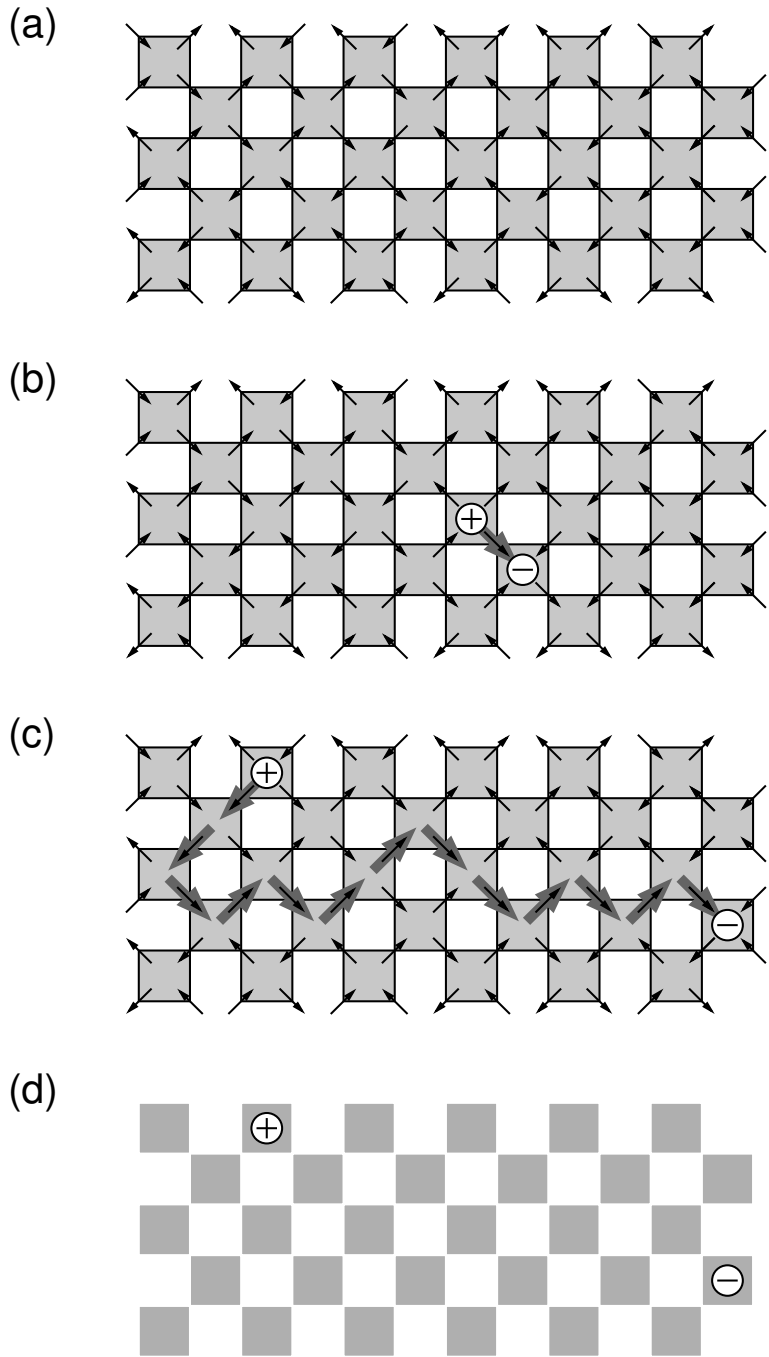

Figure 4: (a) The spin ice problem illustrated on a two-dimensional lattice with each square plaquette exhibiting the 2-in, 2-out arrangement of spins. (b) Reversing a single spin results in two monopoles, which can (c) move independently (as a result of flipping further spins, shown in dark grey). (d) The description of the system at a higher (less cluttered) level is then only in terms of the monopoles and the 'background' spins then become part of the vacuum.